\documentclass{ws-mpla}

\usepackage{url}
\usepackage{overcite}
\usepackage{epsfig}
\usepackage{amssymb}

\newcommand{\Lam}{\Lambda}
\newcommand{\lam}{\lambda}

\newcommand{\mpc}{{\rm Mpc}}

\newcommand{\cm}{\rm cm}

\newcommand{\kmsmpc}{\rm km\,s^{-1}\,Mpc^{-1}}

\newcommand{\Msun}{\rm M_{\odot}}

\newcommand{\Zsun}{Z_{\odot}}
\newcommand{\hinv}{h^{-1}}
\newcommand{\himpc}{\hinv{\rm\,Mpc}}

\newcommand{\himsun}{\hinv{\Msun}}

\newcommand{\Om}{\Omega_{\rm m}}
\newcommand{\Ol}{\Omega_{\Lam}}
\newcommand{\Ob}{\Omega_{\rm b}}
\newcommand{\OHI}{\Omega_{\rm HI}}
\newcommand{\HI}{H\,{\sc i}}

\newcommand{\CII}{[C\,{\sc ii}]}

\newcommand{\NHI}{N_{\rm HI}}
\newcommand{\Nmax}{N_{\rm max}}
\newcommand{\Nmin}{N_{\rm min}}

\newcommand{\sdla}{\sigma_{\rm DLA}}
\newcommand{\sdlaco}{\sigma_{\rm DLA}^{co}}
\newcommand{\sdlaphys}{\sigma_{\rm DLA}^{\rm phys}}

\newcommand{\Mhalo}{M_{\rm halo}}
\newcommand{\Lbox}{{\rm L_{box}}}

\newcommand{\Sigsfr}{\Sigma_{\rm SFR}}

\newcommand{\dd}{{\rm d}}

\newcommand{\Rab}{R_{AB}}
\newcommand{\hsev}{h_{70}}
\newcommand{\hsevkpc}{h_{70}^{-1}\,{\rm kpc}}
\newcommand{\rdla}{r_{\rm DLA}}
\newcommand{\Adla}{A_{\rm DLA}}
\newcommand{\Lcii}{L_{\rm CII}}
\newcommand{\highz}{high-$z$}
\newcommand{\ltsim}{\lesssim}

\newcommand{\rhoc}{\rho_{\rm crit}}
\newcommand{\Snu}{S_{\nu}}

\newcommand{\apj}{ApJ}
\newcommand{\apjl}{ApJL}
\newcommand{\apjs}{ApJS}
\newcommand{\araa}{ARA\&A}
\newcommand{\pasp}{PASP}
\newcommand{\aap}{A\&A}
\newcommand{\mnras}{MNRAS}

\begin{document}

\markboth{Nagamine}
{DLAs and Galaxy Formation}

%%%%%%%%%%%%%%%%%%%%% Publisher's Area please ignore %%%%%%%%%%%%%%
\catchline{}{}{}{}{}
%%%%%%%%%%%%%%%%%%%%%%%%%%%%%%%%%%%%%%%%%%%%%%%%%%%%%%%%%%%%%%%%%%%

\title{DLAs and Galaxy Formation}

\author{\footnotesize KENTARO NAGAMINE
%\footnote{Typeset names in 8 pt Times Roman, uppercase. 
%Use the footnote to indicate the present or permanent address 
%of the author.}
}

\address{Department of Physics \& Astronomy, 
University of Nevada, Las Vegas, \\
4505 Maryland Pkwy, Box 454002, 
Las Vegas, NV 89154-4002, U.S.A.\\
kn@physics.unlv.edu}

\maketitle

\pub{Received (Day Month Year)}{Revised (Day Month Year)}

\begin{abstract}
Damped Lyman-$\alpha$ systems (DLAs) are useful probes
of star formation and galaxy formation at high-redshift 
(hereafter \highz). 
We study the physical properties of DLAs and their relationship 
to Lyman break galaxies (LBGs) using cosmological hydrodynamic 
simulations based on the concordance $\Lambda$ cold dark matter 
model.  
Fundamental statistics such as global neutral hydrogen (\HI) 
mass density, \HI\ column density distribution function, 
DLA rate-of-incidence and mean halo mass of DLAs are 
reproduced reasonably well by the simulations, but with 
some deviations that need to be understood better in the future. 
We discuss the feedback effects by supernovae and galactic winds
on the DLA distribution. 
We also compute the \CII\ emission from neutral gas in \highz\ 
galaxies, and make predictions for the future observations 
by the {\it Atacama Large Millimeter Array} (ALMA) and 
{\it Space Infrared Telescope for Cosmology and Astrophysics}
(SPICA).
Agreement and disagreement between simulations and observations
are discussed, as well as the future directions of our DLA research.

\end{abstract}

%\ccode{PACS Nos.: include PACS Nos.}

\section{Introduction}

Recent studies suggest that our universe can be 
described well by a cosmological model called the 
concordance $\Lambda$ cold dark matter (CDM) model 
with a matter energy density 
$\Om \equiv \rho_{\rm m} / \rhoc \simeq 0.3$ and 
a dark energy density $\Ol \equiv \rho_{\Lam} / \rhoc \simeq 0.7$.
\cite{Ostriker95, Bahcall99, Spergel03}
Based on this backbone of structure formation, we would like to 
build a self-consistent model of galaxy formation that can 
describe the formation of disk, spheroid, and black holes. 
In particular, we would like to understand how the gas turned into  
stars as a function of cosmic time and environment.

Observations of DLAs provide very unique opportunities for 
that purpose. Because DLAs are observed in absorption,
\footnote{DLAs are historically defined\cite{Wolfe86} to be 
quasar absorption systems with \HI\ column density 
$\NHI > 2\times 10^{20} \cm^{-2}$.
See Ref.~\refcite{Wolfe05} for an extensive review on DLAs.}
they are {\em in principle} an unbiased sample of \HI\ in the universe, 
and provide us with information that are complementary 
to those obtained by direct observations of stellar emission. 
For example, with DLAs we can measure the global \HI\ mass 
density in the universe as a function of redshift.\cite{Lanzetta95}  
It has been estimated that DLAs contain a significant fraction of 
\HI\ gas in the universe at $z\sim 3$, with the amount 
roughly equal to a half of current stellar mass density.
\cite{Storr00, Peroux03, Pro05, Jorgenson06}
Therefore DLAs are considered to be important reservoirs
of neutral gas for star formation in \highz\ universe,
and their study would reveal the physical properties of 
neutral gas before they turn into stars. 
The conversion history of neutral gas into stars 
has been an active area of research over the past ten years,
\cite{Madau96, Steidel99, Nag01a, Her03, Ouchi04a, Hopkins06}
and the connection between DLAs and star-forming galaxies 
such as LBGs\cite{Steidel99} is becoming one of the focus
of current DLA research. 

In recent years, observers have made significant progress 
in constraining the statistical properties of DLAs using 
large samples of \highz\ quasars discovered by  
surveys such as the Sloan Digital Sky Survey (SDSS)\cite{York00}.  
For example, Ref.~\refcite{Pro05} have extracted $\approx 500$ DLAs
from the SDSS Data Release 3 and tightened the constraints on the 
column density distribution function and the rate of incidence. 
The total number of detected DLAs has reached over 1000
as of early 2007 (Prochaska, private communication).
Increased accuracy in these observables set tight 
constraints on the models of galaxy formation.

In turn, theoretical understanding of DLAs remain unsatisfactory. 
Currently there are mainly two competing views on the physical 
origin of DLAs. One idea is that DLAs originate from 
rapidly rotating, massive disk galaxies.\cite{Pro97}  
This model can explain the large velocity widths 
of low-ionization metal absorption lines.
However, Ref.~\refcite{Hae98} used hydrodynamic simulations
to argue that the asymmetric profiles of low-ionization 
absorption lines can also be explained by the rotation, 
random motions, infall, and merging of protogalactic 
gas clumps equally well. 
Somewhat intermediate idea is intersecting multiple disks
\cite{Maller01} or tidal tails observed in merging systems. 

There are a number of DLA studies using so-called semianalytic 
models of galaxy formation\cite{Kau96, Okoshi04}, but 
it would also be desirable to study DLAs using cosmological 
hydrodynamic simulations that directly solve the gas dynamics 
within the framework of $\Lam$CDM universe. 
Ref.~\refcite{NSH04a} \& \refcite{NSH04b} focused on the effects 
of star formation and supernova (SN) feedback on DLAs 
that were largely neglected by the earlier numerical work, 
\cite{Gardner97a, Gardner97b, Gardner01}
and showed that the distribution of DLAs could be significantly 
affected by SN feedback. 
Ref.~\refcite{Razoumov06a} studied the effects of radiative transfer
by post-processing the results of cosmological hydrodynamic 
simulations, but their simulations did not include 
models for star formation and SN feedback. 
It is still difficult to treat the radiative transfer 
dynamically from all galaxies in a large volume of space, 
as well as star formation and feedback by SNe and active galactic 
nuclei in large-scale cosmological hydrodynamic simulations.
This is primarily due to the heavy computational load
required for such a simulation with full treatment 
of all physical processes.
At the moment, we need to limit our calculation to a 
small number of objects in order to treat radiative transfer 
self-consistently with the star formation in simulations. 

In this review article, we summarize our recent numerical 
studies on DLAs using cosmological hydrodynamic simulations.
We discuss the conflicts and agreement between simulations 
and observations, 
and the future directions of theoretical research on DLAs.

%%%%%%%%%%%%%%%%%%%%%%%%%%%%%%%%%%%%%%%%%%%%%%%%%%

\section{Simulating DLAs}

We use cosmological smoothed particle hydrodynamics (SPH) 
code GADGET-2 (Springel 2005) for our study. 
The code solves the hydrodynamics in a Lagrangian fashion
by allowing the gas particles to cluster in high density regions,
thereby providing higher resolution in galaxy-forming regions. 
The gravity is solved by a Tree-particle-mesh method that 
uses a tree algorithm for the short-range force and 
particle-mesh technique for the long-range force. 
It also adopts the entropy-conservative formulation\cite{SH02} 
to remedy the overcooling problem, which previous generation 
of cosmological SPH codes suffered from.  
The code includes radiative cooling, heating by a uniform 
UV background radiation of a modified Haardt \& Madau 
spectrum\cite{Haardt96, Katz96a, Dave99},  
star formation, supernova feedback, galactic wind\cite{SH03b}, 
and a sub-resolution multiphase model of interstellar medium 
(ISM)\cite{SH03a}. 

We employ a series of simulations with different resolution, 
box sizes (from $\Lbox=4\himpc$ to $100\himpc$) and 
feedback strengths to study their effects. 
One of the focus of our study is to examine the effects of 
galactic wind feedback on the distribution of DLAs.
In this article, we summarize our results by referring to 
two cases: `no wind' feedback and `strong wind' feedback, 
corresponding to the wind speed of $v_w = 0$ and 484\,km\,s$^{-1}$.
See Ref.~\refcite{NSH04b} for more details of our simulations. 
All of our simulations are based on a concordance 
$\Lambda$CDM cosmology with cosmological parameters 
$(\Om, \Ol, \Ob,\sigma_8, h) = 
(0.3, 0.7, 0.04, 0.9, 0.7)$, where $h = H_0 / (100\,\kmsmpc)$.  

%%%%%%%%%%%%%%%%%%%%%%%%%%%%%%%%%%%%%%%%%%%%%%%%%%

\section{Fundamental DLA Statistics}

\subsection{H\,{\small I} Column Density Distribution}
\label{sec:fnh}

The \HI\ column density distribution function $f(\NHI)$ is 
the most fundamental statistic for any \HI\ quasar absorption systems. 
It counts the number of columns in bins of 
[$\NHI, \NHI+d\NHI$] per unit absorption distance $dX$, 
i.e., $dN = f(\NHI)d\NHI dX$, where $dX \equiv 
\frac{H_0}{c} (1+z)^3\,c\,dt = \frac{H_0}{c}\,(1+z)^2\, dr$.  
The quantity $dN$ can also be interpreted as an area covering 
fraction in the sky along the line element $c\,dt$. 

\begin{figure}[t]
\begin{center}
\resizebox{6.1cm}{!}{\includegraphics{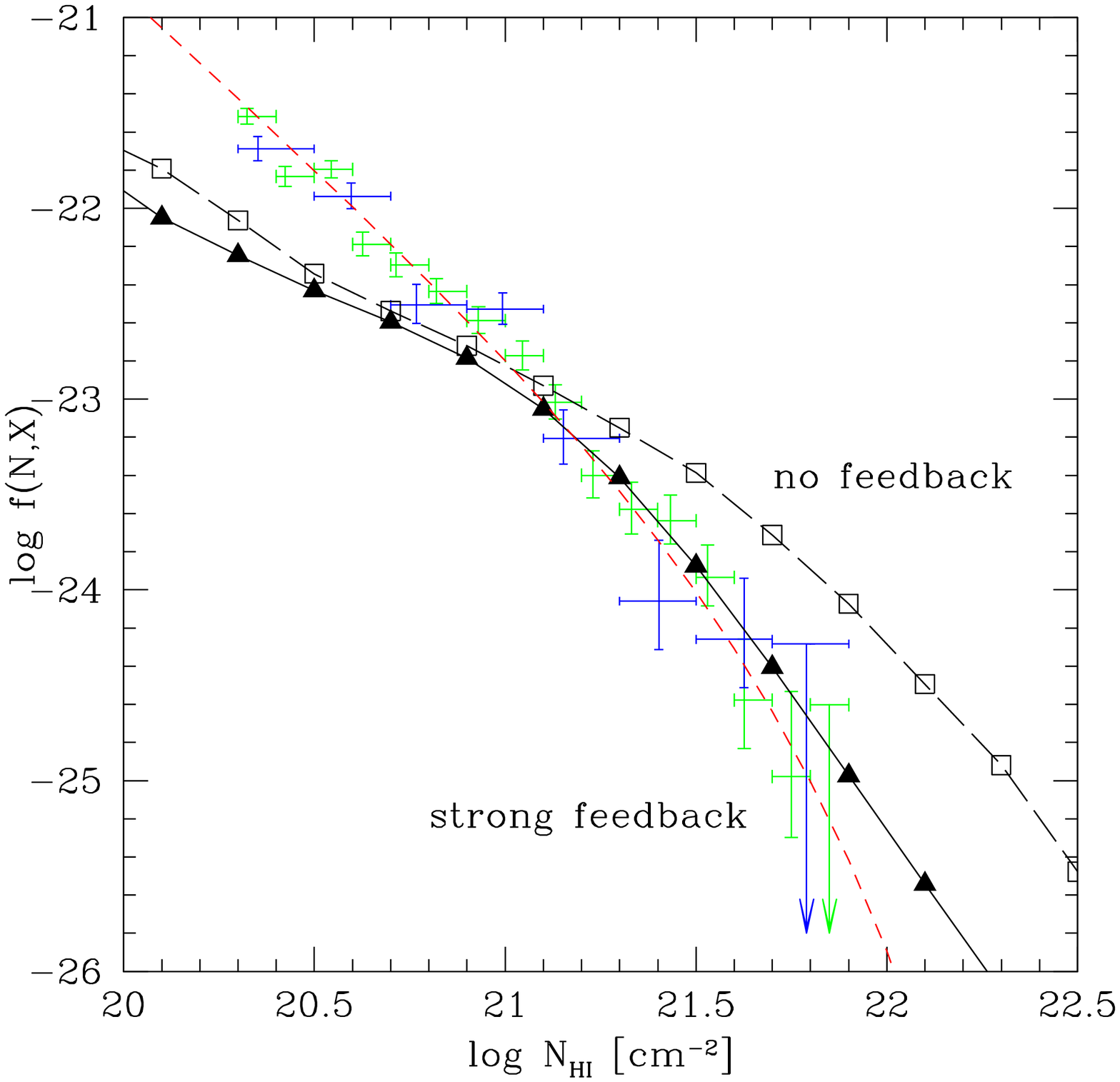}}
\resizebox{6.45cm}{!}{\includegraphics{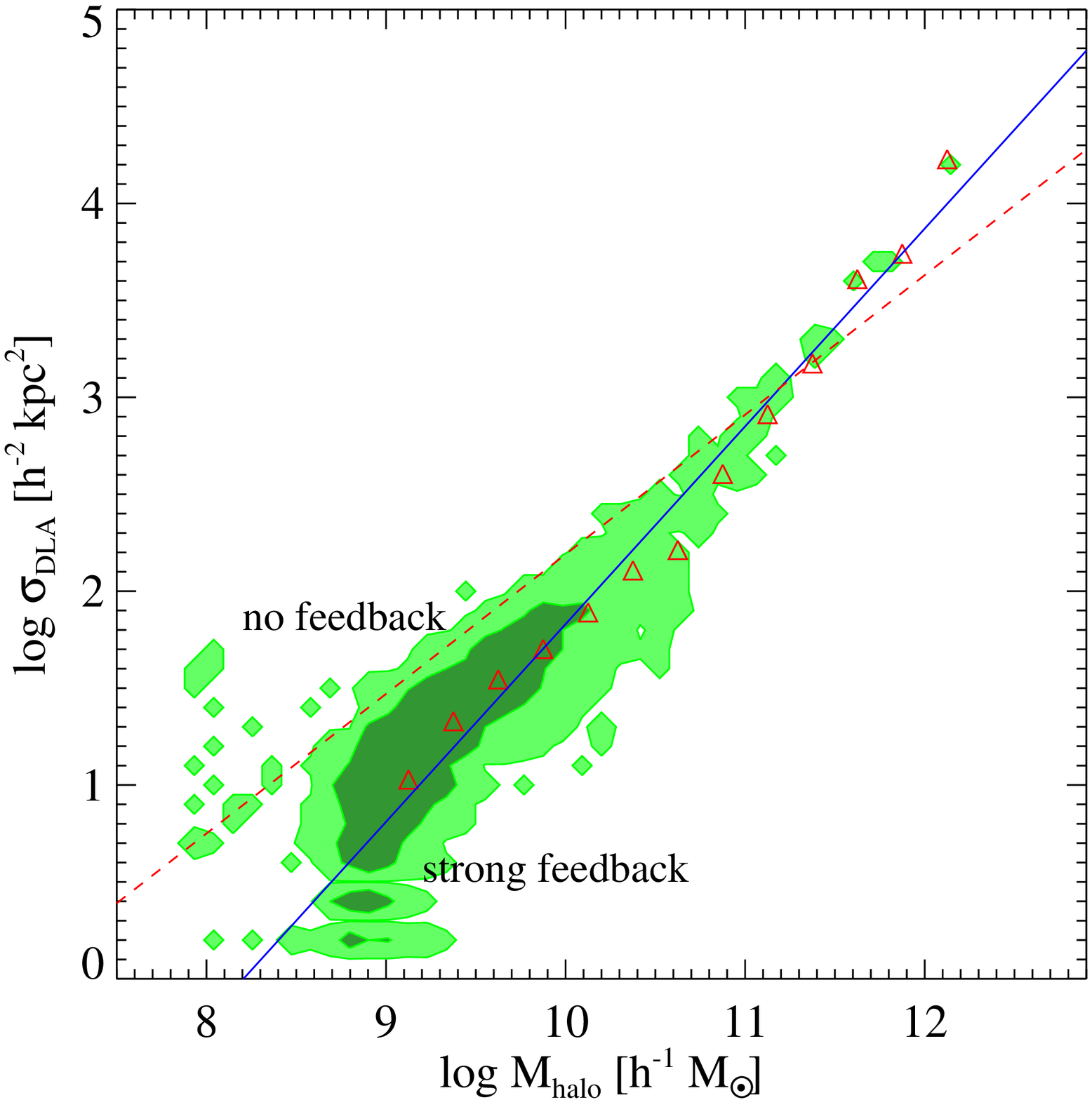}} \\
\caption{{\it Left}: Column density distribution function $f(\NHI)$
for the runs with no wind (open squares) and 
strong wind feedback (solid triangles).
The red short-dashed line is the gamma function fit to the 
observed data points\cite{Pro05}.
{\it Right}: Comoving DLA cross section as a function of halo mass. 
The shaded contours show the distribution of halos for the run
with strong wind feedback.  The open triangle symbols are the 
median points in each mass bin, and the solid line is the 
least-square fit to the median points. 
The dashed line is the same fit for the run with no wind feedback. 
}
\label{fig:column}
\end{center}
\end{figure}

Ref.~\refcite{NSH04a} examined $f(\NHI)$ in a series of
cosmological SPH simulations with different box sizes and 
resolution.
The left panel of Fig.~\ref{fig:column} shows 
the results for runs with no wind and strong wind feedback, 
together with the observed data points.\cite{Pro05}
At high column densities ($\log \NHI > 21$), the agreement 
between the strong feedback run and observed data is quite good.  
The run with no wind feedback significantly overpredicts $f(\NHI)$. 
This is because the galactic wind feedback reduces 
the number of columns with large $\NHI$ by ejecting the gas 
from star-forming regions and heat the \HI\ gas. 
This agreement at $\log \NHI>21$ for the strong feedback run
is encouraging and shows the effectiveness of our current
wind feedback model. 
One might speculate that the \HI\ in the overpredicted columns
in the no feedback run could turn into molecular hydrogen, therefore
the apparent disagreement between the simulation and the data points
is not a serious problem.  However, Ref.~\refcite{Zwaan06} showed
that the H$_2$ column density distribution forms 
a natural extension to $f(\NHI)$ for local galaxies, and 
that the connection point between the two distribution is at
$(\log \NHI, \log f(\NHI))\approx (22, -26)$.
Since the degree of overprediction in the no feedback run 
is more than a factor of 2 at $\log \NHI \ltsim 22$, 
it seems difficult to hide all of the overpredicted \HI\ in the 
form of H$_2$ if similar atomic and molecular
physics hold in \highz\ galaxies as in local galaxies. 

Our simulations underpredict the number of columns
with $20 < \log \NHI <21$ regardless of the strength of 
wind feedback. 
Physical reasons for this underestimate is not clear at this point.
Several test runs show that this feature does not change very much
with varying threshold density or time-scale for star formation, 
or varying wind feedback strength. 
Any combination of the following could be the cause
for this underestimate:
inadequate resolution, angular momentum transfer problem, 
inadequate cooling due to lack of metal cooling in the current code, 
too simplistic models for star formation and SN feedback, 
or poor treatment of radiative transfer. 
We plan to address these issues in our future work.

%%%%%%%%%%%%%%%%%%%%%%%%%%%%%%%%%%%%%%%%%%%%%%%%%%

\subsection{Neutral Gas Mass Density}
\label{sec:h1density}

For a given observed path length $c \Delta t$, 
the \HI\ mass density within the survey volume
can be estimated as follows (in proper units):
\begin{equation}
\rho_{\rm HI} = \frac{m_{\rm H}}{c\Delta t} \int_{\Nmin}^{\Nmax} 
\NHI f(\NHI) d\NHI \Delta X 
= \frac{H_0 m_{\rm H} (1+z)^3}{c} \int_{\Nmin}^{\Nmax} \NHI f(\NHI) d\NHI,  
\label{eq:rhoh1}
\end{equation}
where $m_H$ is the mass of a hydrogen atom, and 
$[\Nmin, \Nmax]$ is the observed range of $\NHI$. 
If the observed range of $\NHI$ is broad enough
and covers up to sufficiently high $\NHI$ values,  
then one can expect that the calculated quantity is 
representative of the true value in the universe, 
and the mass density parameter of neutral gas 
can be estimated as 
$\Omega_{\rm neutral} = \mu \rho_{\rm HI} / \rhoc$, 
 where $\mu \approx 1.3$ is inserted to take helium
into account. 
It is known observationally\cite{Storr00, Peroux03, Pro05} 
that DLAs dominate the \HI\ mass density at $z\sim 3$, 
with the amount roughly equal to half of the present-day
stellar mass density. This suggests that DLAs are
important reservoirs of neutral gas for star formation
in \highz\ galaxies, and that the neutral gas in DLAs
at $z=3$ are converted into stars between $z=3$ and 
$z=0$.  
Ref.~\refcite{NSH04a} showed that the observed
estimates of $\Omega_{\rm neutral}$ can be bracketed
by the simulation with no wind and strong wind feedback.
It can also be shown that high-$\NHI$ systems dominate
$\OHI$, therefore it is important to include a fair 
number of high-$\NHI$ systems in the sample to obtain 
an accurate fit to $f(\NHI)$ and estimate $\OHI$ reliably.

%%%%%%%%%%%%%%%%%%%%%%%%%%%%%%%%%%%%%%%%%%%%%%%%%% 

\subsection{DLA Cross Section vs. Halo Mass}
\label{sec:sigma}

It is useful to quantify the cross section of DLAs ($\sdla$) 
as a function of dark matter halo mass, because we have a 
theoretical framework to compute the halo mass function.
\cite{Press74, Sheth99} 
Observers can identify DLAs only in the quasar sight-lines, 
but there could be multiple neutral gas clouds 
within the same halo that do not cross the quasar sight-line. 
Therefore it is theoretically more plausible to characterize 
the occurrence of DLAs as a cross section rather than 
the number of clouds in each halo. 

Ref.~\refcite{NSH04a} estimated the relationship between 
the total DLA cross section $\sdlaco$ (in units of 
comoving $h^{-2}$\,kpc$^2$) and the dark matter halo mass 
(in units of $\himsun$) at $z=3$ as
\begin{equation}
\log\,\sdlaco = \alpha\, (\log\,\Mhalo - 12) + \beta,
\label{eq:sigma}
\end{equation}
with slopes $\alpha = 0.7 - 1.0$ and normalization 
$\beta = 3.9 - 4.2$, depending on the numerical resolution 
and the strength of galactic wind feedback.
The slope $\alpha$ is always positive and massive halos
have larger DLA cross sections, but they are more scarce 
compared to less massive halos. 
Two important features are:
(1) As the strength of galactic wind feedback increases, 
the slope $\alpha$ become steeper while the normalization $\beta$ 
remains roughly constant. 
This is because a stronger wind reduces the amount of 
neutral gas in low-mass halos at a higher rate by ejecting 
the gas out of the potential well of the halo.  
(2) As the numerical resolution is improved, both the slope 
and the normalization increase. 
This is because with higher resolution, simulations
can resolve higher densities, and star formation in 
low-mass halos can be described better.  
As a result the neutral gas content 
is decreased due to winds. 
On the other hand, a lower resolution run misses the 
early generation of halos and the neutral gas in them, 
resulting in lower $\OHI$ at \highz.

Earlier numerical work\cite{Gardner97a, Gardner97b, Gardner01} 
largely neglected the effects of 
star formation and wind feedback, therefore they often 
overestimated the DLA cross sections in low-mass halos, 
resulting in higher value of predicted rate-of-incidence 
(see \S~\ref{sec:rate}).  
In order to match their results with observed data, 
they introduced a minimum halo mass that DLAs populate. 
Our simulations show that star formation and feedback
affect the DLA distribution, therefore, we need to study 
their effects further with better resolution and improved 
models of star formation and feedback. 

%%%%%%%%%%%%%%%%%%%%%%%%%%%%%%%%%%%%%%%%%%%%%%%%%%

\subsection{DLA Rate-of-Incidence}
\label{sec:rate}

The probability of finding a DLA for a given observed path-length 
can be expressed as a rate-of-incidence. 
Theoretically, we can compute the cumulative rate-of-incidence 
of DLAs per unit redshift above certain minimum halo mass $M$ as
\begin{equation}
\frac{{\rm d}N_{\rm DLA}}{{\rm d}z}(>M) = \frac{{\rm d}r}{{\rm d}z} 
\int_M^{\infty} n_{\rm dm}(M')~\sdlaco(M')\,{\rm d}M',
\label{eq:abundance}
\end{equation}
where $n_{\rm dm}(M)$ is the dark matter halo mass function
and $\dd r/\dd z = c/H(z)$ with $H(z)=H_0\sqrt{\Om(1+z)^3+\Ol}$ 
for a flat universe.  In order to carry out this integral, 
Eq.~(\ref{eq:sigma}) can be used. 

\begin{figure}[t]
\begin{center}
\resizebox{6.2cm}{!}{\includegraphics{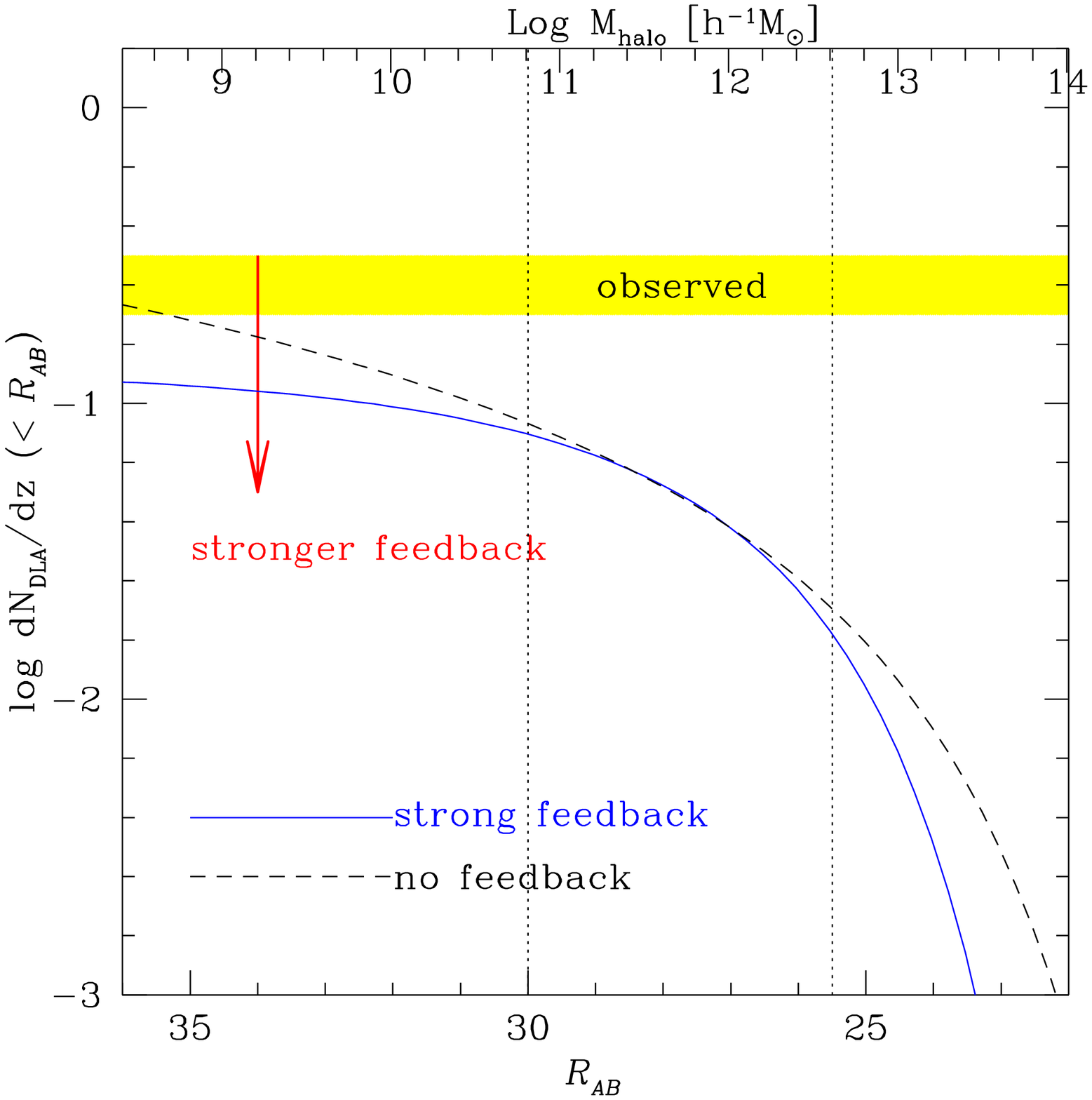}}
%\hspace{0.13cm}
\resizebox{6.2cm}{!}{\includegraphics{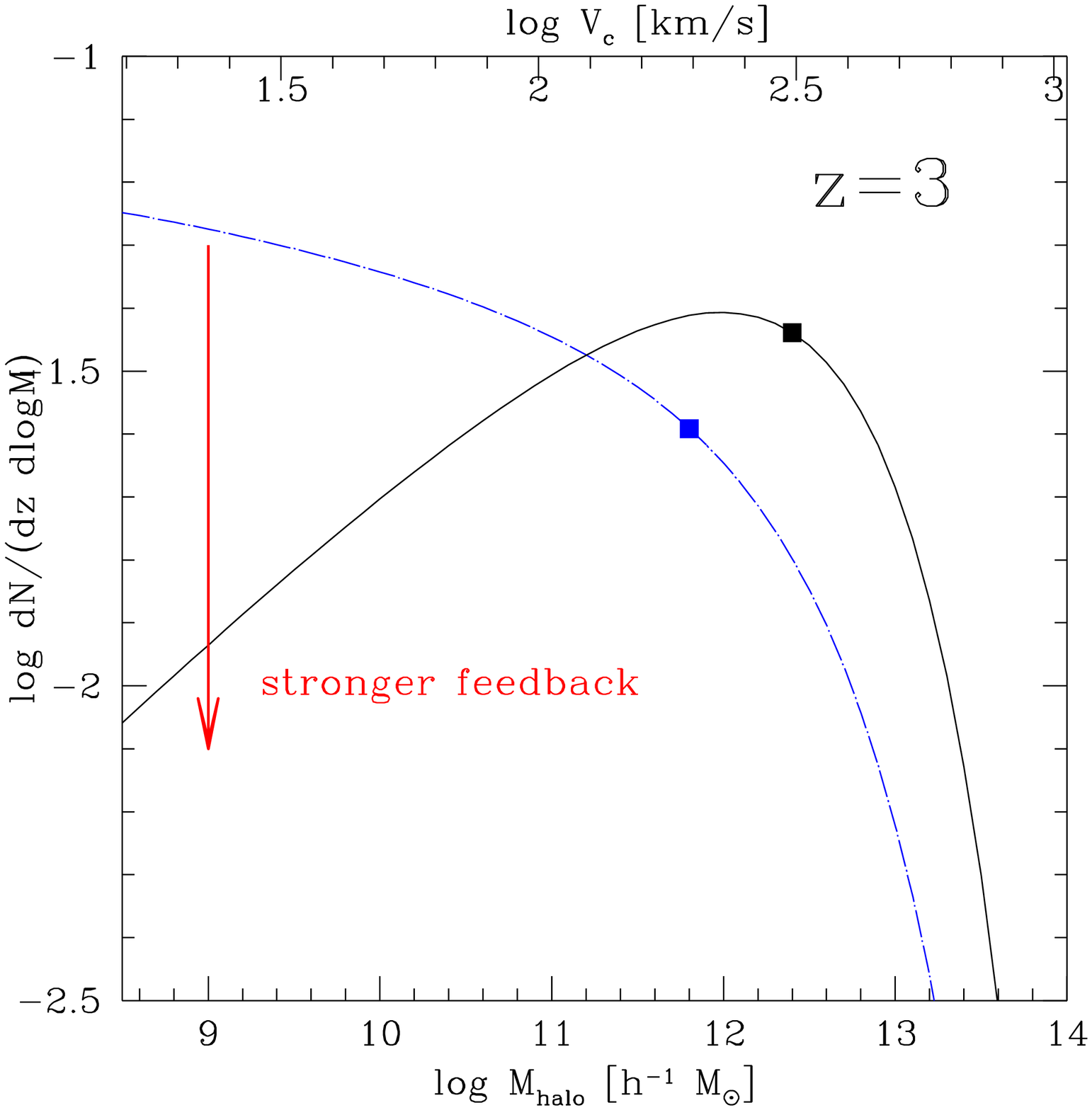}} \\
%\vspace{0.1cm}
\caption{{\it Left}: Cumulative DLA rate-of-incidence for 
all galaxies above certain magnitude limit. 
An observational estimate\cite{Pro05} is shown by the hatched band. 
{\it Right}: Probability distribution function of 
DLA rate-of-incidence as a function of dark matter halo mass.  
As the feedback strength increases, contribution from 
low-mass halos decreases, and the distribution shifts toward 
higher masses with a peak emerging at 
$\Mhalo \simeq 10^{12}\himsun$. 
}
\label{fig:rate}
\end{center}
\end{figure}

The left panel of Fig.~\ref{fig:rate} shows that
the run with no wind feedback agrees with the observed rate 
if we sample the halos down to $\Mhalo \simeq 10^{8.5}\himsun$. 
The run with strong feedback underestimates the total 
DLA rate-of-incidence by about 0.3 dex due to 
the underestimate of $f(\NHI)$ at $\log \NHI < 21$.  
Based on the results of hydro simulations, 
Ref.~\refcite{Nag07} derived the mean scaling relation 
between halo mass and galaxy $R$-band AB magnitude as
\begin{equation}
\Rab = -2.5\log \Mhalo + C - 5\log \hsev, 
\label{eq:mag}
\end{equation}
where $C=55.03$ (no feedback) and $57.03$ (strong feedback).   
This relation was used to convert the halo mass into 
apparent $\Rab$ magnitude in the left panel of Fig.~\ref{fig:rate}.  
The arrows in the figure indicate that the wind feedback
reduces the contribution from low-mass halos to 
the DLA rate-of-incidence. 

One can also calculate the differential probability distribution 
of DLAs as a function of halo mass as
\begin{equation}
\frac{\dd N_{\rm DLA}}{\dd z\, \dd\log M} = 
\frac{\dd r}{\dd z} [\, M\, n(M)\, \ln(10)\,]\, \sdlaco(M).
\label{eq:rate}
\end{equation}
See Ref.~\refcite{Nag07} for the derivation of 
Eq.~(\ref{eq:abundance}) and (\ref{eq:rate}).
Since stronger wind feedback suppresses the contribution 
from low-mass halos to the total DLA rate-of-incidence, 
the distribution shifts toward higher mass halos as the 
feedback strength is increased, and eventually
a peak in the distribution appears at $\Mhalo \simeq 10^{12}\himsun$ 
as shown in the right panel of Fig.~\ref{fig:rate}. 
This mass-scale is interesting, because recent 
observational studies\cite{Cooke06a, Cooke06b} 
of cross-correlation between DLAs and LBGs 
suggested similar characteristic halo masses for DLAs.
Ref.~\refcite{Bouche05} also estimated the mean halo mass of DLAs
using numerical simulations (see Ref.~\refcite{Nag07} for a
detailed comparison of our results with theirs.).

%%%%%%%%%%%%%%%%%%%%%%%%%%%%%%%%%%%%%%%%%%%%%%%%%%

\subsection{Impact Parameter Distribution}

\begin{figure}[t]
\begin{center}
\resizebox{6.1cm}{!}{\includegraphics{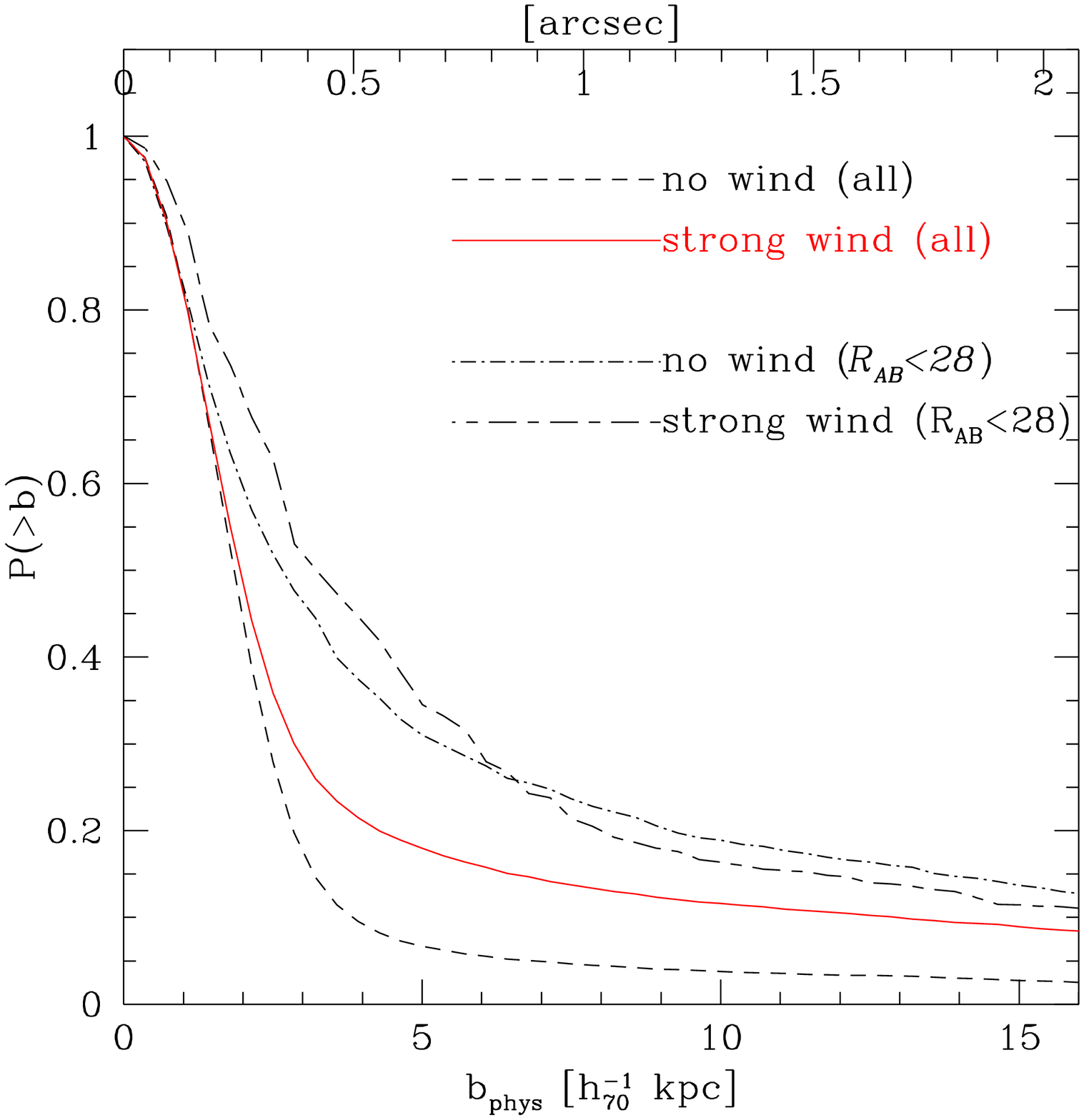}}
\hspace{0.1cm}
\resizebox{6.1cm}{!}{\includegraphics{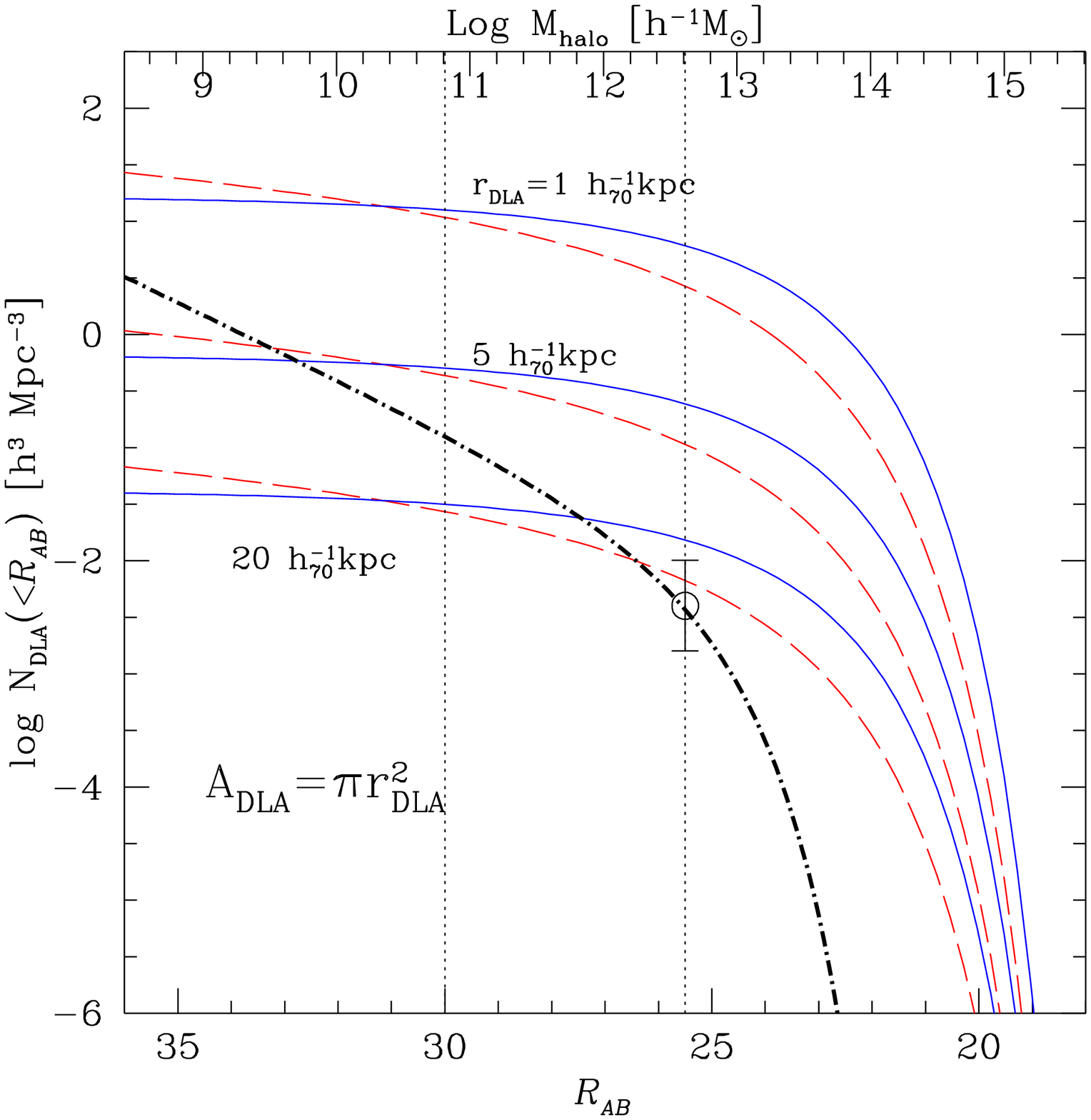}} \\%
\caption{
{\it Left}: Normalized impact parameter distribution for 
runs with and without galactic wind feedback.  
The results with limited search for the nearest galaxy 
to $\Rab<28$ mag are also shown. 
{\it Right}: Cumulative comoving number density of DLAs 
as a function of apparent $\Rab$ magnitude for three 
different values of assumed DLA physical radius 
$\rdla = 1\,\hsevkpc$, 5\,$\hsevkpc$, and 20\,$\hsevkpc$.
The covering area of each DLA is $A_{\rm DLA} = \pi \rdla^2$. 
For each value of $\rdla$, results from no feedback run 
(long-dashed line) and  strong feedback run (solid line) 
are shown. The thick dot-dashed line is the cumulative 
comoving number density of LBGs obtained by integrating 
the observed luminosity function\cite{Ade00}.
The data point at $\Rab=25.5$ shows the 
observed comoving number density of LBGs 
$N_{\rm LBG} =  4\times 10^{-3} h^{-3}\,\mpc^{-3}$. 
}
\label{fig:impact}
\end{center}
\end{figure}

Another important way to characterize the spatial distribution 
of DLAs is to examine the distribution of impact parameter 
``$b$'' between DLAs and the nearest galaxy. 
For a given DLA sight-line in the simulation, we search for the 
nearest galaxy on the projected plane, and determine 
the $b$-parameter for each DLA. 
As the left panel of Fig.~\ref{fig:impact} shows, 
the $b$-parameter distribution is quite narrow 
in the current simulations, and over 80\% of 
DLAs have a galaxy within physical $5\hsevkpc$. 
This large fraction of DLAs with small $b$-parameter is
due to the large number of low-mass halos in a CDM universe
and DLAs associated with them.  As the Figs.~$1-3$ in 
Ref.~\refcite{NSH04b} show, DLAs in lower mass halos 
are very compact with physical sizes of a few kpc, 
and coincide with star-forming regions very well. 
One might consider that this is plausible given the fact that the 
physical radii of LBGs are $2-3$\,kpc\cite{Law07},  
however, DLAs could also originate from extended \HI\ clouds 
observed as, e.g., tidal tails of merging systems. %%\cite{} 
We do not seem to have such extended DLAs associated with 
low-mass halos in our current simulations.

If we limit the search for nearby galaxies to those 
brighter than $\Rab = 28$\,mag, 
then the distribution becomes broader and roughly 
80\% of DLAs have a galaxy within physical $10\hsevkpc$.  
If we increase the strength of galactic wind feedback, 
then we reduce the fraction of DLAs in low-mass halos,
and a larger fraction of DLAs will be in more massive halos
that are farther away, resulting in a broader $b$-parameter 
distribution. 
It is not clear at this point whether these conclusions from 
the current simulations are consistent with current observations
given the limited sample.\cite{Moller04, Chen05}
When a larger sample of DLA host galaxies is constructed, 
the comparison will become more interesting.

We can also constrain the spatial distribution of 
DLAs using the cross-correlation function\cite{Gawiser01}   
between DLAs and galaxies.
Ref.~\refcite{Cooke06a} computed the cross correlation function 
using 211 LBGs and 11 DLAs, and showed that there
is a strong correlation between the two population, 
comparable to the LBG auto-correlation. 
This suggests that many DLAs are physically related to LBGs 
and quite possibly in the same dark matter halos as LBGs. 
We will present the DLA-LBG cross-correlation function 
in our simulations in the future.

%%%%%%%%%%%%%%%%%%%%%%%%%%%%%%%%%%%%%%%%%%%%%%%%%%

\subsection{Number Density of DLAs}

It would be useful to quantify the number density of 
DLAs similarly to the galaxy number density. 
However doing so is not easy both observationally and 
in numerical simulations, because DLA gas clouds 
are not detected in emission and they are often 
extended, connected, and exhibit very complicated 
morphologies as shown in Fig.~1 of Ref.~\refcite{NSH04b}.

Theoretically, it is possible to calculate the mean number density 
of DLA clouds as follows.  Assuming that the average 
physical covering area of each DLA gas cloud is fixed with  
$\Adla = \pi\,\rdla^2$, then the cumulative number density of DLAs 
above a certain halo mass $M$ is 
\begin{equation}
N_{\rm DLA}(>M) = \int_M^{\infty} dM\, n_{\rm dm}(M)\, \frac{\sdlaphys(M)}{\Adla},
\label{eq:ndla}
\end{equation}
where $n_{\rm dm}(M)$ is the dark matter halo mass function, 
and $\sdlaphys(M)$ is the physical DLA cross section
as a function of halo mass. 
Note that $\sdla$ is the {\it total}
DLA cross section of each dark matter halo; 
in other words, if there are 100 DLA clouds in a massive halo, 
then this halo has a total DLA cross section of 
$\sdlaphys = 100\, \Adla$. 
Then, using Eq.~(\ref{eq:mag}) and (\ref{eq:ndla}), 
we calculate the cumulative comoving number density of 
DLAs as a function of apparent $\Rab$ magnitude 
$N_{\rm DLA}(<\Rab)$ as shown in 
the right panel of Fig.~\ref{fig:impact}. 
This figure shows that the number density of DLA clouds 
is higher than that of LBGs if their physical radius is 
smaller than $\rdla = 20 \hsevkpc$. 
Based on a similar calculation, Ref.~\refcite{Schaye01a}
suggested possible relationship between outflows from LBGs and 
DLAs, although this idea is yet to be confirmed observationally.

%%%%%%%%%%%%%%%%%%%%%%%%%%%%%%%%%%%%%%%%%%%%%%%%%%

\subsection{Metallicity, $\NHI$, and Star Formation Rate}

Measurements of gas metallicity in DLAs and galaxies 
provide useful information on star formation and 
chemical enrichment history. 
The mean metallicity of DLAs is known\cite{Pettini04} 
to be $\approx (1/30)\Zsun$, 
and that of LBGs to be $\approx (1/3)\Zsun$. 

\begin{figure}[t]
\begin{center}
\resizebox{6.1cm}{!}{\includegraphics{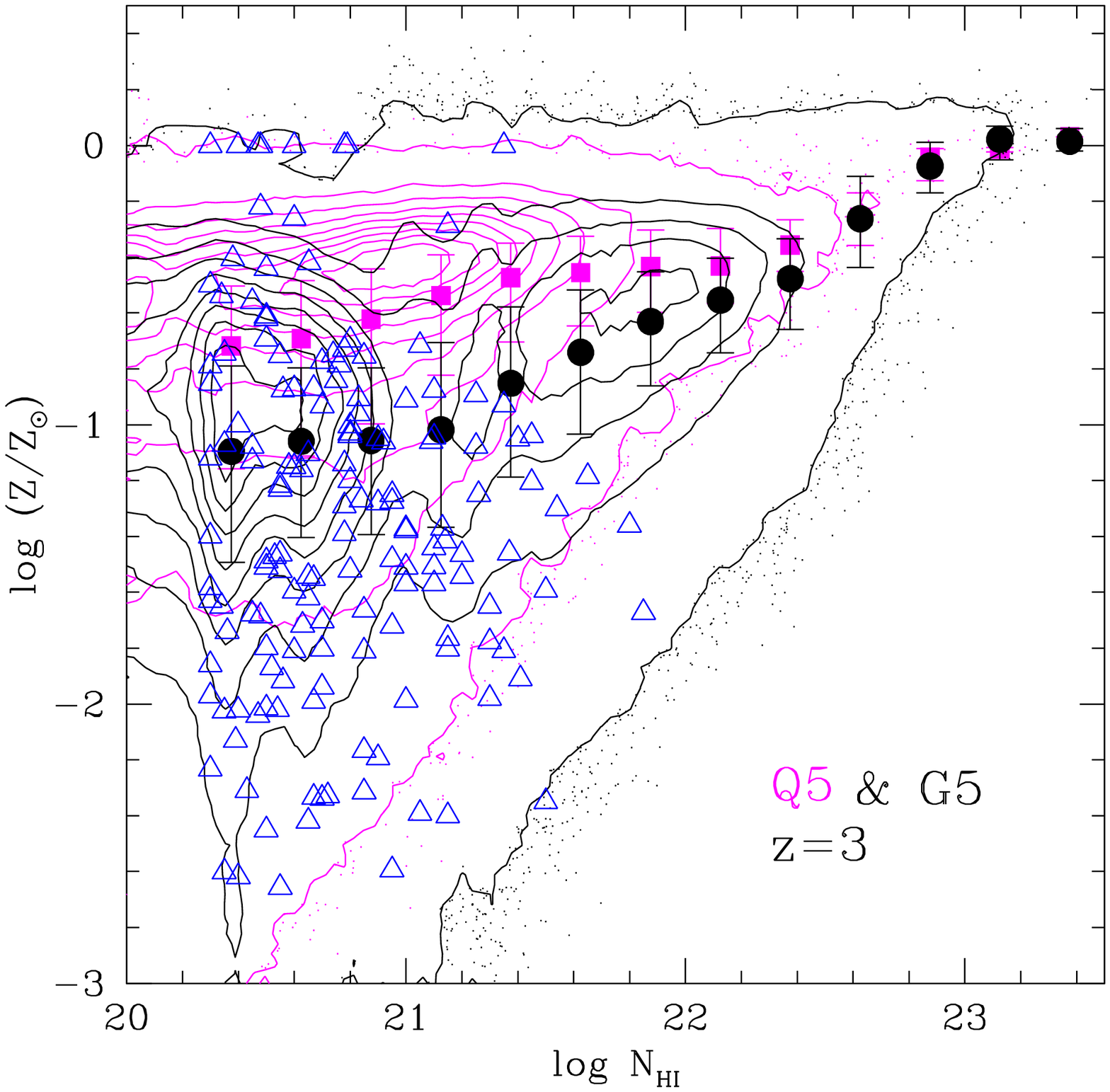}}
\hspace{0.1cm}
\resizebox{6.1cm}{!}{\includegraphics{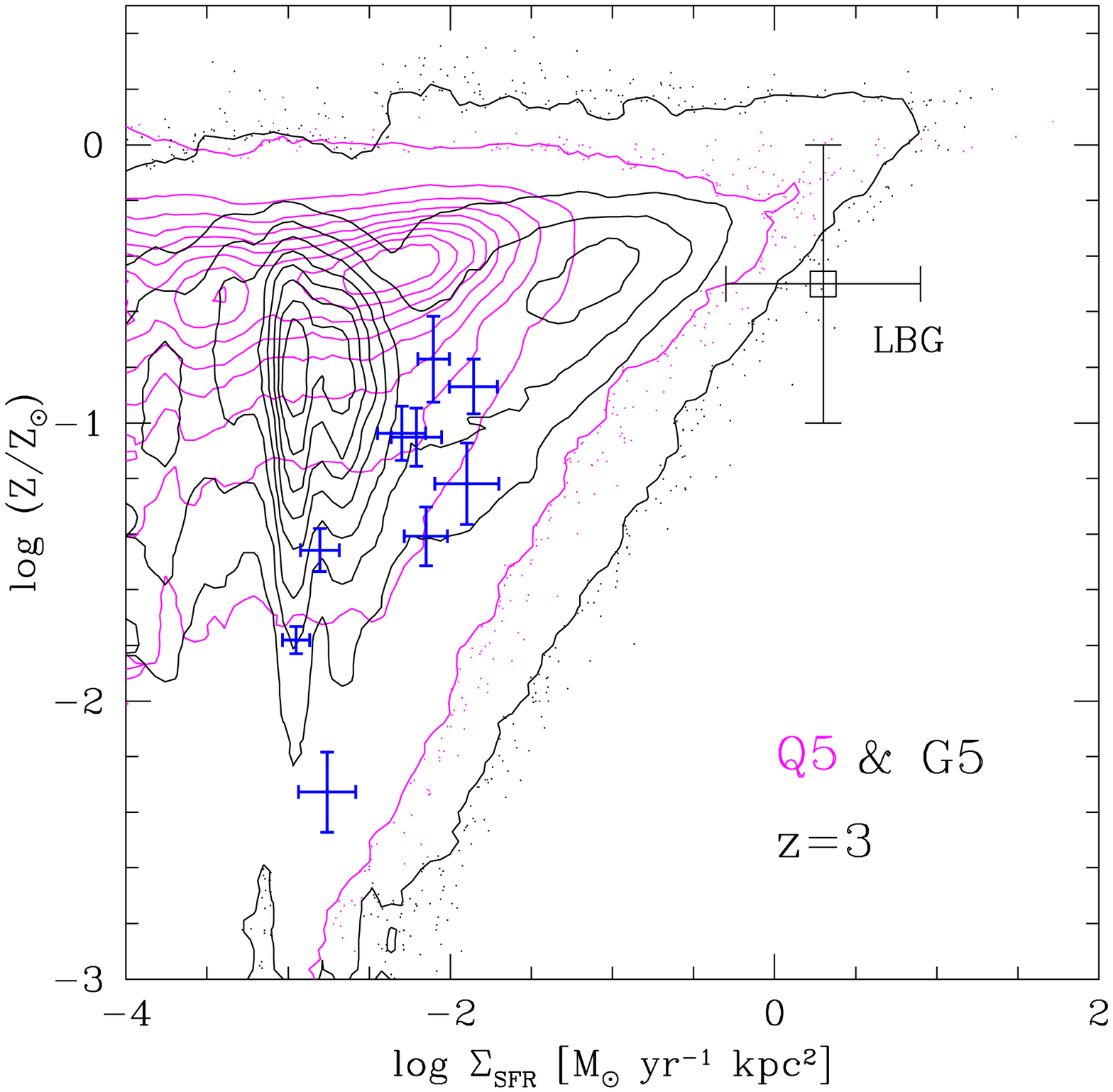}} \\
\caption{
{\it Left:} Distribution of DLA columns on the plane of 
\HI\ column density vs. metallicity are shown for
the Q5 run ($\Lbox=10\,\himpc$; magenta contour) and the 
G5 run ($\Lbox=100\,\himpc$; black contour)
at $z=3$ with the median points for each $\log \NHI$ bin. 
The blue open triangles are observed data\cite{Pro07}.
{\it Right:} Same as the left panel for the plane of 
projected star formation rate vs. metallicity. 
The skewed contour shape at $\log \Sigsfr \approx -3$
for the G5 run is a resolution effect. 
The blue data points are observational estimates
by the C\,{\sc ii}$^*$ technique\cite{Wolfe03b}, 
and the open square symbol represents LBGs roughly. 
}
\label{fig:metal}
\end{center}
\end{figure}

In Figure~\ref{fig:metal}, we show the distribution of 
gas metallicity as functions of $\NHI$ and projected 
star formation rate $\Sigsfr$ for the two strong wind 
feedback runs (Q5 and G5 run) at $z=3$.
The G5 run has a broader distribution than the Q5 run 
because of a larger box size ($\Lbox=100\,\himpc$ for the 
G5 and $10\,\himpc$ for the Q5 run), but lower median 
metallicity at lower $\NHI$ values due to lower resolution. 
In the left panel, we see that there are no observed DLAs\cite{Pro07} 
with $\log\NHI > 22.0$ and $\log (Z/\Zsun)>-1.0$, while such 
sight-lines exist in our simulations. 
These columns with high $\NHI$ and high metallicity are 
probably associated with central cores of LBGs that 
are actively forming stars. 
The occurrence of such columns are very rare as can be
seen from the rapidly declining $f(\NHI)$ function, and 
they have very small cross-sections. 
There are other ideas to explain the absence of DLAs at 
$\log\NHI > 22.0$. One popular idea is the extinction of 
background quasars by dust\cite{Fall93, Vladilo05}. 
Another idea is that the conversion of \HI\ to H$_2$ molecule
determines the maximum value of $\NHI$.\cite{Schaye01b}
It is not fully clear at this point which of the above three
explanations (or any combinations of them) is the correct one.  
At $\log\NHI < 22.0$, the Q5 run might be overpredicting
the median metallicity of DLAs compared to observations. 

The right panel of Fig.~\ref{fig:metal} shows a similar 
distribution as the left panel. 
This is because $\Sigsfr$ and $\NHI$ are correlated with 
each other strongly (i.e., $\Sigsfr \propto \NHI^{1.4}$, 
the Kennicutt Law\cite{Kennicutt98}; 
see also Fig.~8 of Ref.~\refcite{NSH04b}), 
and metallicity also correlates with star formation positively.
The metallicity for the columns with $\log \Sigsfr \gtrsim 0$
in the simulation might be slightly too high 
compared to the observational estimates of LBGs.
The observational estimates by the C\,{\sc ii}$^*$ 
technique\cite{Wolfe03b} (blue data points; only positive 
detections are shown) nicely fall onto the simulated
contour. 

%%%%%%%%%%%%%%%%%%%%%%%%%%%%%%%%%%%%%%%%%%%%%%%%%%

\section{Predicting the [C\,{\small II}] Line Emission}
\label{sec:cii}

The \CII\ emission line at $\lambda=157.74\,\mu$m originates
from the $^2P_{3/2} \rightarrow ^2P_{1/2}$ 
fine structure transition of C$^+$. 
This emission line is often the brightest emission 
in the spectrum of galaxies, and it is suggested to be 
the dominant coolant of diffuse ISM at temperatures 
$T \le 5000$ K.\cite{Dalgarno72, Tielens85, Wolfire95, Lehner04} 
Therefore, it has been suggested that the \CII\ emission 
is potentially another method for finding cold 
neutral gas in \highz\ universe.\cite{Petrosian69, Loeb93} 

Locally, \CII\ emission has been detected from 
both 1) dense star-forming gas irradiated by UV radiation 
from young stars\cite{Russell80, Crawford85, Stacey91, Carral94}
and 2) extended, diffuse cold components of ISM in quiescent late-type
galaxies\cite{Madden93, Leech99, Malhotra01, Contursi02}.
At \highz, there is only a few tentative 
detections\cite{Maiolino05, Iono06} of 
\CII\ emission from quasars at $z=6.42$ and $z=4.7$, 
but none for more normal
(i.e., not quasars or active galactic nuclei) star-forming 
galaxies such as LBGs.  Detections of \CII\ emission from 
LBGs, DRGs\cite{Franx03, Dokkum04}, and star-forming 
BzK\cite{Daddi04b} galaxies might become possible 
in the near future by ALMA and SPICA.

We might be able to set important constraints on 
galaxy formation models using \CII\ detections. 
In particular interferometric maps of \CII\ emission 
would reveal the spatial dimensions of neutral gas
that is responsible for DLAs, 
a property that has so far eluded detection. 
By combining these measurements with the velocity widths 
of the \CII\ emission lines, it will be possible to 
infer the dark matter halo masses of the sources, 
and the \CII\ luminosity would tell us the heating rate 
of the gas.

\begin{figure}
\begin{center}
\resizebox{6.1cm}{!}{\includegraphics{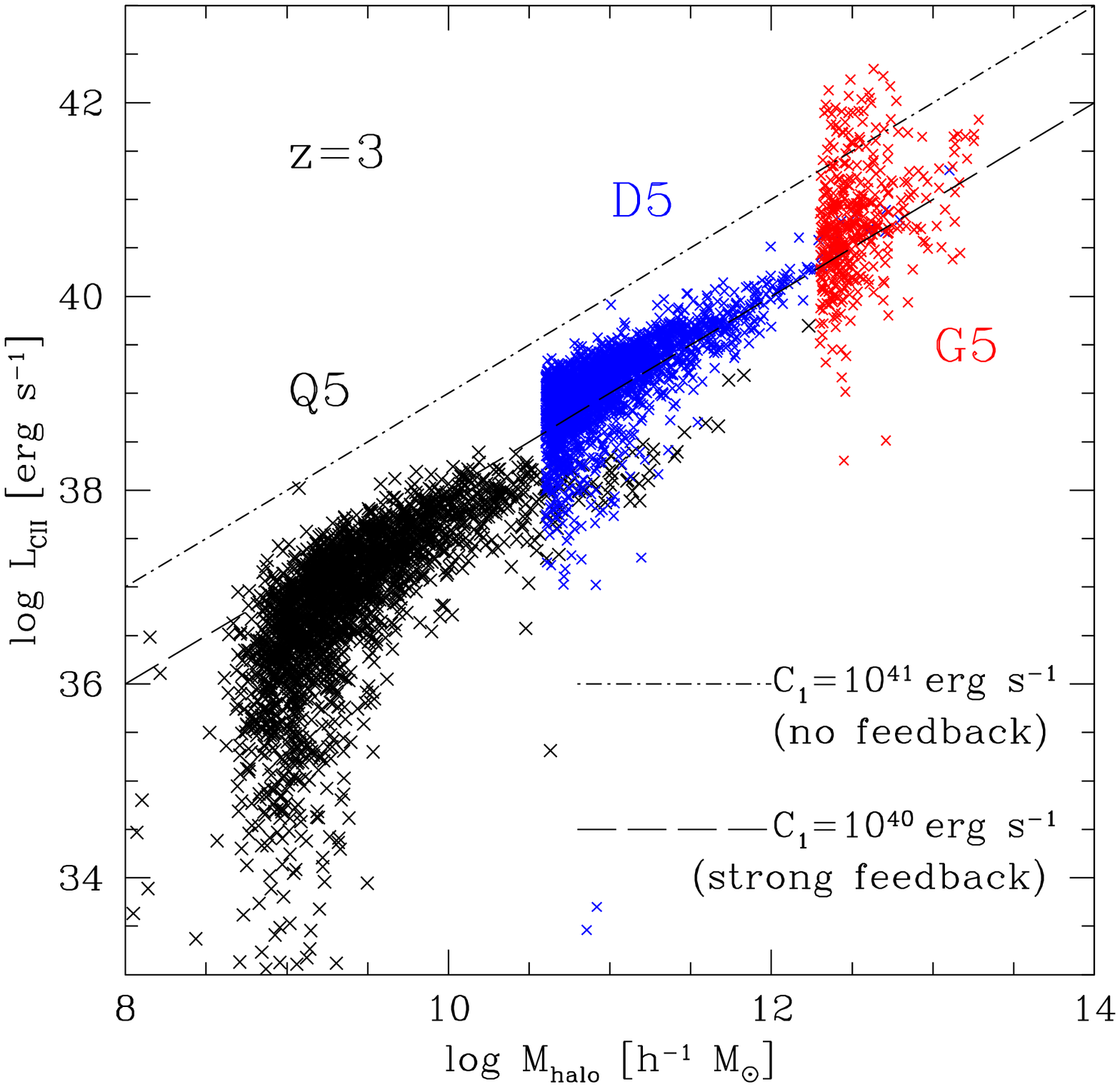}}
\hspace{0.1cm}
\resizebox{6.1cm}{!}{\includegraphics{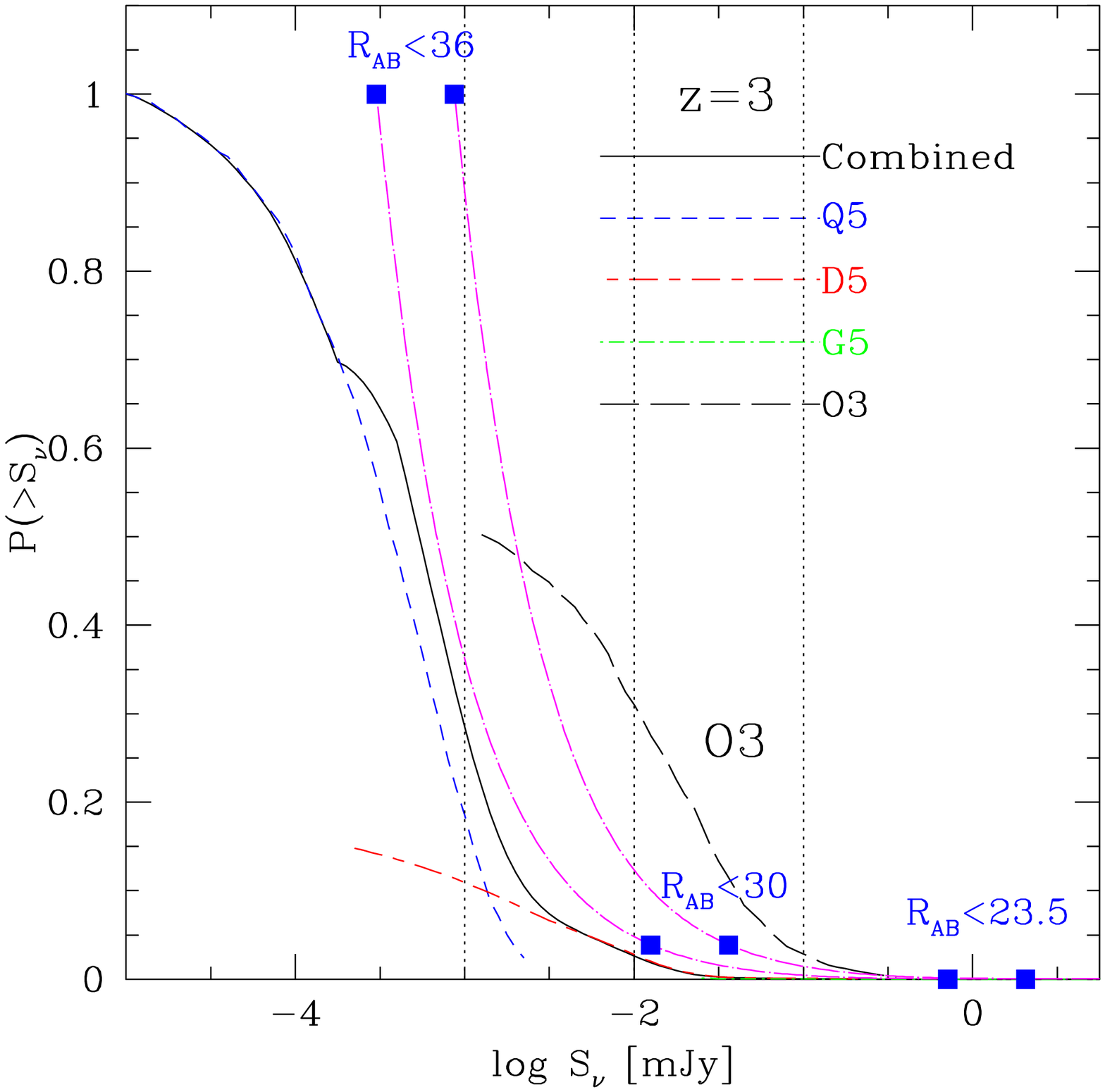}} \\%
%\vspace{0.1cm}
\caption{{\it Left}: \CII\ luminosity as a function of 
halo mass at $z=3$. The three sets of points are
all from strong wind feedback runs: Q5, D5, and G6 runs
with box sizes $\Lbox=10, 34$, and 100$\himpc$, respectively.
{\it Right}: Cumulative probability distribution function 
of \CII\ sources at $z=3$ as a function of flux density. 
The result from the run with no wind feedback 
(the O3 run with $\Lbox=10\himpc$, long-dashed line) is also shown. 
The two magenta dot-dashed lines are obtained\cite{Nag06b} 
by integrating the observed luminosity function\cite{Ade00} 
of LBGs.
The three vertical dotted lines indicate the flux density limits
of $\Snu=0.1, 0.01,$ and 0.001\,mJy.
the diagram. 
}
\label{fig:cii}
\end{center}
\end{figure}

Ref.~\refcite{Nag06b} attempted to make a realistic estimate
of \CII\ emission from \highz\ galaxies by 
coupling the analytic model of Ref.~\refcite{Wolfe03a} 
and cosmological hydrodynamic simulations. 
We first used the star formation rate and metallicity of gas 
in our simulations to estimate the amount of cold neutral 
medium (CNM; typically $T\sim 100$\,K and $n\sim 10$\,cm$^{-3}$) 
in each halo, and then computed the \CII\ luminosity 
by multiplying the \CII\ luminosity per \HI\ atom
estimated by the analytic model to the CNM mass.  
As shown in the left panel of Fig.~\ref{fig:cii}, we find
that the \CII\ luminosity can be described roughly by 
\begin{equation}
\Lcii = C_1 \left( \frac{\Mhalo}{10^{12}\,\himsun} \right),
\label{eq:Lcii}
\end{equation}
where $C_1 = 10^{41}$ \,erg\,s$^{-1}$ (no wind) 
and $10^{40}$\,erg\,s$^{-1}$ (strong wind feedback). 
The strong wind feedback reduces the amount of CNM
by ejecting the gas from star-forming regions, 
resulting in lower \CII\ luminosity by a factor of $\sim 10$
compared to the no feedback run.

The right panel of Fig.~\ref{fig:cii} shows the probability 
distribution function of \CII\ sources at $z=3$
as a function of \CII\ flux density. 
For the runs with strong feedback,  
we see that only a few percent of all the sources 
would have \CII\ flux brighter than 1\,mJy.  
Most of the \CII\ sources will be very faint, 
low-mass galaxies with $\Snu < 0.1$\,mJy 
due to the large number of low-mass halos in a 
$\Lambda$CDM model.  
The run with no wind feedback gives more optimistic 
result in terms of the fraction of bright sources, 
suggesting that $\sim 30$\% of \CII\ sources have 
$\Snu > 0.01$\,mJy at $z=3$.  
These predictions can be tested by ALMA and SPICA in the future. 
Even though the \CII\ observation of normal \highz\ galaxies 
is currently difficult, the reward of \CII\ detection from such 
sources would be quite significant, because such 
measurements could directly constrain the amount of 
CNM and the properties of ambient ISM, 
which is otherwise difficult to achieve.

The crude calculation by Ref.~\refcite{Nag06b} was 
satisfactory as a first step, but we need to treat 
the radiation from local sources and the ionization of gas 
more accurately.  To this end, we plan to use the 
photoionization code CLOUDY\cite{Ferland98} and refine
our \CII\ calculations in the future. 

%%%%%%%%%%%%%%%%%%%%%%%%%%%%%%%%%%%%%%%%%%%%%%%%%%

\section{Conclusions \& Future Research}

In this article, we reviewed our DLA research 
using cosmological SPH simulations.  
We achieved some successes in explaining 
the global DLA statistics crudely, 
but there are details that need to be studied further. 
In particular, our simulations underpredict $f(\NHI)$ at
$\log \NHI < 21.0$.  
To cure this problem, we need to make improvements on 
the following points:
1) increase mass and spatial resolution, 
2) treat radiative transfer from local radiation sources 
and photoionization of gas more accurately, 
3) improve the models for star formation and feedback. 
Also, our simulations are probably still suffering from 
the angular momentum transfer\cite{Kaufmann07} problem, 
which we hope to remedy as we work on the above three points. 

The possible connection between DLAs and LBGs is 
quite interesting. Ref.~\refcite{Wolfe03a} detected the 
C\,{\sc ii}$^*$ $\lam$1335.7 absorption in about 
a half of randomly selected DLAs, and showed that 
local heat sources (such as LBGs) are required
to match the heating rate inferred from the \CII\
cooling rate.  Their work and Ref.~\refcite{Cooke06a} 
suggest close connection between DLAs and LBGs. 
It would be natural for the two population to be
closely related to each other, considering the 
fact that DLAs are important reservoirs of 
neutral gas for star formation at \highz, 
and that LBGs contribute significantly 
to the total star formation rate density in \highz\ universe.

Just like the multi-wavelength observations of galaxies can 
reveal different properties of galaxies, 
we can also tackle the mysteries of DLAs using 
different techniques including \HI\ absorption, 
C\,{\sc ii}$^*$ and other metal line absorption, 
and \CII\ emission. 
Through continuous collaboration between observation and 
theory, we hope to clarify the physical nature of 
DLAs in the future.

%\begin{figure}[th]
%\centerline{\psfig{file=mplaf1.eps,width=2.0in}}
%\vspace*{8pt}
%\caption{A schematic illustration of dissociative recombination. The
%direct mechanism, 4$m^2_\pi$ is initiated when the
%molecular ion $S_{L}$ captures an electron with kinetic 
%energy.\protect\label{fig1}}
%\end{figure}

%\begin{table}[h]
%\tbl{Comparison of acoustic for frequencies for piston-cylinder problem.}
%{\begin{tabular}{@{}cccc@{}} \toprule
%Piston mass & Analytical frequency & TRIA6-$S_1$ model &
%\% Error \\
%& (Rad/s) & (Rad/s) \\ 
%\colrule
%1.0\hphantom{00} & \hphantom{0}281.0 & \hphantom{0}280.81 & 0.07 \\
%0.1\hphantom{00} & \hphantom{0}876.0 & \hphantom{0}875.74 & 0.03 \\
%0.01\hphantom{0} & 2441.0 & 2441.0\hphantom{0} & 0.0\hphantom{0} \\
%0.001 & 4130.0 & 4129.3\hphantom{0} & 0.16\\ \botrule
%\end{tabular}}
%\end{table}

\section*{Acknowledgments}

The author would like to thank his collaborators, Art Wolfe, 
Lars Hernquist and Volker Springel for our DLA research.
Simulations and analyses for this paper were performed at 
the Center for Parallel Astrophysical Computing at 
Harvard-Smithsonian Center for Astrophysics and
the UNLV Cosmology Computing Cluster.

%\section*{References}

%\bibliographystyle{ws-mpla}
%\bibliography{ken,apj-jour}

\begin{thebibliography}{10}

\bibitem{Ostriker95}
J.~P. Ostriker and P.~J. Steinhardt, {\em Nature} {\bf 377},   600 (1995).

\bibitem{Bahcall99}
N.~A. Bahcall, J.~P. Ostriker, S.~Perlmutter and P.~Steinhardt, {\em Science}
  {\bf 284},   1481 (1999).

\bibitem{Spergel03}
D.~Spergel, L.~Verde, H.~V. Peiris, E.~Komatsu, M.~R. Nolta, C.~L. Bennett,
  M.~Halpern, G.~Hinshaw {\em et~al.}, {\em ApJS} {\bf 148},   175 (2003).

\bibitem{Wolfe86}
A.~M. {Wolfe}, D.~A. {Turnshek}, H.~E. {Smith} and R.~D. {Cohen}, {\em ApJS}
  {\bf 61}, 249 (1986).

\bibitem{Wolfe05}
A.~M. Wolfe, E.~Gawiser and J.~X. Prochaska, {\em ARA\&A} {\bf 43},   861
  (2005).

\bibitem{Lanzetta95}
K.~M. Lanzetta, A.~M. Wolfe and D.~A. Turnshek, {\em ApJ} {\bf 440},   435
  (1995).

\bibitem{Storr00}
L.~J. {Storrie-Lombardi} and A.~M. {Wolfe}, {\em ApJ} {\bf 543}, 552 (2000).

\bibitem{Peroux03}
C.~P\'{e}roux, R.~G. McMahon, L.~J. Storrie-Lombardi and M.~J. Irwin, {\em
  MNRAS} {\bf 346},   1103 (2003).

\bibitem{Pro05}
J.~X. {Prochaska}, S.~{Herbert-Fort} and A.~M. {Wolfe}, {\em ApJ} {\bf 635},
  123 (2005).

\bibitem{Jorgenson06}
R.~A. {Jorgenson}, A.~M. {Wolfe}, J.~X. {Prochaska}, L.~{Lu}, J.~C. {Howk},
  J.~{Cooke}, E.~{Gawiser} and D.~M. {Gelino}, {\em ApJ} {\bf 646},   730
  (2006).

\bibitem{Madau96}
P.~Madau, H.~C. Ferguson, E.~D. Dickinson, M.~Giavalisco, C.~C. Steidel and
  A.~Fruchter, {\em MNRAS} {\bf 283},   1388 (1996).

\bibitem{Steidel99}
C.~C. Steidel, K.~L. Adelberger, M.~Giavalisco, M.~Dickinson and M.~Pettini,
  {\em ApJ} {\bf 519},  ~1 (1999).

\bibitem{Nag01a}
K.~Nagamine, M.~Fukugita, R.~Cen and J.~P. Ostriker, {\em ApJ} {\bf 558},   497
  (2001).

\bibitem{Her03}
L.~Hernquist and V.~Springel, {\em MNRAS} {\bf 341},   1253 (2003).

\bibitem{Ouchi04a}
M.~Ouchi, K.~Shimasaku, H.~Furusawa, M.~Miyazaki, M.~Doi, M.~Hamabe,
  T.~Hayashino, M.~Kimura {\em et~al.}, {\em ApJ} {\bf 611},   660 (2004a).

\bibitem{Hopkins06}
A.~M. {Hopkins} and J.~F. {Beacom}, {\em ApJ} {\bf 651}, 142 (2006).

\bibitem{York00}
D.~G. York {\em et~al.}, {\em AJ} {\bf 120},   1579 (2000).

\bibitem{Pro97}
J.~X. Prochaska and A.~M. Wolfe, {\em ApJ} {\bf 487},  ~73 (1997).

\bibitem{Hae98}
M.~Haehnelt, M.~Steinmetz and M.~Rauch, {\em ApJ} {\bf 495},   647 (1998).

\bibitem{Maller01}
A.~H. {Maller}, J.~X. {Prochaska}, R.~S. {Somerville} and J.~R. {Primack}, {\em
  MNRAS} {\bf 326}, 1475 (2001).

\bibitem{Kau96}
G.~Kauffmann, {\em MNRAS} {\bf 281},   475 (1996).

\bibitem{Okoshi04}
K.~Okoshi, M.~Nagashima, N.~Gouda and S.~Yoshioka, {\em ApJ} {\bf 603},  ~12
  (2004).

\bibitem{NSH04a}
K.~Nagamine, V.~Springel and L.~Hernquist, {\em MNRAS} {\bf 348},   421 (2004).

\bibitem{NSH04b}
K.~Nagamine, V.~Springel and L.~Hernquist, {\em MNRAS} {\bf 348},   435 (2004).

\bibitem{Gardner97a}
J.~Gardner, N.~Katz, L.~Hernquist and D.~H. Weinberg, {\em ApJ} {\bf 484},  ~31
  (1997a).

\bibitem{Gardner97b}
J.~Gardner, N.~Katz, D.~H. Weinberg and L.~Hernquist, {\em ApJ} {\bf 486},  ~42
  (1997b).

\bibitem{Gardner01}
J.~Gardner, N.~Katz, L.~Hernquist and D.~H. Weinberg, {\em ApJ} {\bf 559},
  131 (2001).

\bibitem{Razoumov06a}
A.~O. {Razoumov}, M.~L. {Norman}, J.~X. {Prochaska} and A.~M. {Wolfe}, {\em
  ApJ} {\bf 645}, 55 (2006a).

\bibitem{SH02}
V.~Springel and L.~Hernquist, {\em MNRAS} {\bf 333},   649 (2002).

\bibitem{Haardt96}
F.~Haardt and P.~Madau, {\em ApJ} {\bf 461},  ~20 (1996).

\bibitem{Katz96a}
N.~Katz, D.~H. Weinberg and L.~Hernquist, {\em ApJS} {\bf 105},  ~19 (1996).

\bibitem{Dave99}
R.~Dav\'{e}, L.~Hernquist, N.~Katz and D.~H. Weinberg, {\em ApJ} {\bf 511},
  521 (1999).

\bibitem{SH03b}
V.~Springel and L.~Hernquist, {\em MNRAS} {\bf 339},   312 (2003).

\bibitem{SH03a}
V.~Springel and L.~Hernquist, {\em MNRAS} {\bf 339},   289 (2003).

\bibitem{Zwaan06}
M.~A. {Zwaan} and J.~X. {Prochaska}, {\em \apj} {\bf 643}, 675 (2006).

\bibitem{Press74}
W.~H. {Press} and P.~{Schechter}, {\em ApJ} {\bf 187}, 425 (1974).

\bibitem{Sheth99}
R.~K. Sheth and G.~Tormen, {\em MNRAS} {\bf 308},   119 (1999).

\bibitem{Nag07}
K.~{Nagamine}, A.~M. {Wolfe}, L.~{Hernquist} and V.~{Springel}, {\em \apj} {\bf
  660}, 945 (2007).

\bibitem{Cooke06a}
J.~{Cooke}, A.~M. {Wolfe}, E.~{Gawiser} and J.~X. {Prochaska}, {\em ApJL} {\bf
  636}, L9 (2006).

\bibitem{Cooke06b}
J.~{Cooke}, A.~M. {Wolfe}, E.~{Gawiser} and J.~X. {Prochaska}, {\em \apj} {\bf
  652}, 994 (2006).

\bibitem{Bouche05}
N.~Bouche, J.~P. Gardner, D.~H. Weinberg, R.~Dav\'{e} and J.~D. Lowenthal, {\em
  ApJ} {\bf 628},  ~89 (2005).

\bibitem{Law07}
D.~R. {Law}, C.~C. {Steidel}, D.~K. {Erb}, M.~{Pettini}, N.~A. {Reddy}, A.~E.
  {Shapley}, K.~L. {Adelberger} and D.~J. {Simenc}, {\em \apj} {\bf 656}, 1
  (2007).

\bibitem{Moller04}
P.~M{\o}ller, J.~U. Fynbo and S.~M. Fall, {\em A\&A} {\bf 422},   L33 (2004).

\bibitem{Chen05}
H.-W. {Chen}, J.~X. {Prochaska}, B.~J. {Weiner}, J.~S. {Mulchaey} and G.~M.
  {Williger}, {\em ApJL} {\bf 629}, L25 (2005).

\bibitem{Gawiser01}
E.~Gawiser, A.~M. Wolfe, J.~X. Prochaska, K.~M. Lanzetta, N.~Yahata and
  A.~Quirrenbach, {\em ApJ} {\bf 562},   628 (2001).

\bibitem{Ade00}
K.~L. Adelberger and C.~C. Steidel, {\em ApJ} {\bf 544},   218 (2000).

\bibitem{Schaye01a}
J.~{Schaye}, {\em \apjl} {\bf 559}, L1 (2001).

\bibitem{Pettini04}
M.~{Pettini}, {Element abundances through the cosmic ages}, in {\em
  Cosmochemistry. The melting pot of the elements\/},  eds. C.~{Esteban},
  R.~{Garc{\'{\i}}a L{\'o}pez}, A.~{Herrero} and F.~a.-p. {S{\'a}nchez}2004.

\bibitem{Pro07}
J.~X. {Prochaska}, A.~M. {Wolfe}, J.~C. {Howk}, E.~{Gawiser}, S.~M. {Burles}
  and J.~{Cooke}, {\em \apjs} {\bf 171}, 29 (2007).

\bibitem{Fall93}
S.~M. {Fall} and Y.~C. {Pei}, {\em \apj} {\bf 402}, 479 (1993).

\bibitem{Vladilo05}
G.~{Vladilo} and C.~{P{\'e}roux}, {\em \aap} {\bf 444}, 461 (2005).

\bibitem{Schaye01b}
J.~{Schaye}, {\em \apjl} {\bf 562}, L95 (2001).

\bibitem{Kennicutt98}
R.~C.~J. Kennicutt, {\em ApJ} {\bf 498},   541 (1998).

\bibitem{Wolfe03b}
A.~M. Wolfe, E.~Gawiser and J.~X. Prochaska, {\em ApJ} {\bf 593},   235 (2003).

\bibitem{Dalgarno72}
A.~{Dalgarno} and R.~A. {McCray}, {\em \araa} {\bf 10}, 375 (1972).

\bibitem{Tielens85}
A.~G.~G.~M. {Tielens} and D.~{Hollenbach}, {\em ApJ} {\bf 291}, 722 (1985).

\bibitem{Wolfire95}
M.~G. {Wolfire}, D.~{Hollenbach}, C.~F. {McKee}, A.~G.~G.~M. {Tielens} and
  E.~L.~O. {Bakes}, {\em ApJ} {\bf 443}, 152 (1995).

\bibitem{Lehner04}
N.~Lehner, B.~P. Wakker and B.~D. Savage, {\em ApJ} {\bf 615},   767 (2004).

\bibitem{Petrosian69}
V.~{Petrosian}, J.~N. {Bahcall} and E.~E. {Salpeter}, {\em ApJL} {\bf 155},
  L57 (1969).

\bibitem{Loeb93}
A.~{Loeb}, {\em ApJL} {\bf 404}, L37 (1993).

\bibitem{Russell80}
R.~W. {Russell}, G.~{Melnick}, G.~E. {Gull} and M.~{Harwit}, {\em ApJL} {\bf
  240}, L99 (1980).

\bibitem{Crawford85}
M.~K. {Crawford}, R.~{Genzel}, C.~H. {Townes} and D.~M. {Watson}, {\em ApJ}
  {\bf 291}, 755 (1985).

\bibitem{Stacey91}
G.~J. Stacey, N.~Geis, R.~G. ad~J.~B.~Lugten, A.~Poglitsch, A.~Sternberg and
  C.~H. Townes, {\em ApJ} {\bf 373},   423 (1991).

\bibitem{Carral94}
P.~{Carral}, D.~J. {Hollenbach}, S.~D. {Lord}, S.~W.~J. {Colgan}, M.~R. {Haas},
  R.~H. {Rubin} and E.~F. {Erickson}, {\em ApJ} {\bf 423}, 223 (1994).

\bibitem{Madden93}
S.~C. {Madden}, N.~{Geis}, R.~{Genzel}, F.~{Herrmann}, J.~{Jackson},
  A.~{Poglitsch}, G.~J. {Stacey} and C.~H. {Townes}, {\em ApJ} {\bf 407}, 579
  (1993).

\bibitem{Leech99}
K.~J. {Leech}, H.~J. {V{\"o}lk}, I.~{Heinrichsen}, H.~{Hippelein},
  L.~{Metcalfe}, D.~{Pierini}, C.~C. {Popescu}, R.~J. {Tuffs} and C.~{Xu}, {\em
  MNRAS} {\bf 310}, 317 (1999).

\bibitem{Malhotra01}
S.~Malhotra, M.~J. Kaufman, D.~Hollenbach, G.~Helou, R.~H. Rubin, J.~Brauher,
  D.~Dale, N.~Y. Lu {\em et~al.}, {\em ApJ} {\bf 561},   766 (2001).

\bibitem{Contursi02}
A.~Contursi, M.~J. Kaufman, G.~Helou, D.~J. Hollenbach {\em et~al.}, {\em AJ}
  {\bf 124},   751 (2002).

\bibitem{Maiolino05}
R.~{Maiolino}, P.~{Cox}, P.~{Caselli}, A.~{Beelen}, F.~{Bertoldi}, C.~L.
  {Carilli}, M.~J. {Kaufman}, K.~M. {Menten} {\em et~al.}, {\em \aap} {\bf
  440}, L51 (2005).

\bibitem{Iono06}
D.~{Iono}, M.~S. {Yun}, M.~{Elvis}, A.~B. {Peck}, P.~T.~P. {Ho}, D.~J.
  {Wilner}, T.~R. {Hunter}, S.~{Matsushita} and S.~{Muller}, {\em \apjl} {\bf
  645}, L97 (2006).

\bibitem{Franx03}
M.~Franx, I.~Labbe, G.~Rudnick, P.~G. van Dokkum, E.~Daddi, S.~F\"{o}rster,
  M.~Natascha, A.~Moorwood {\em et~al.}, {\em ApJ} {\bf 587},   L79 (2003).

\bibitem{Dokkum04}
P.~van Dokkum, M.~Franx, N.~F. Schreiber, G.~Illingworth, E.~Daddi, K.~K.
  Knudsen, I.~Labbe, A.~Moorwood {\em et~al.}, {\em ApJ} {\bf 611},   703
  (2004).

\bibitem{Daddi04b}
E.~{Daddi}, A.~{Cimatti}, A.~{Renzini}, A.~{Fontana}, M.~{Mignoli},
  L.~{Pozzetti}, P.~{Tozzi} and G.~{Zamorani}, {\em ApJ} {\bf 617}, 746 (2004).

\bibitem{Nag06b}
K.~{Nagamine}, A.~M. {Wolfe} and L.~{Hernquist}, {\em ApJ} {\bf 647}, 60
  (2006).

\bibitem{Wolfe03a}
A.~M. Wolfe, J.~X. Prochaska and E.~Gawiser, {\em ApJ} {\bf 593},   215 (2003).

\bibitem{Ferland98}
G.~J. {Ferland}, K.~T. {Korista}, D.~A. {Verner}, J.~W. {Ferguson}, J.~B.
  {Kingdon} and E.~M. {Verner}, {\em \pasp} {\bf 110}, 761 (1998).

\bibitem{Kaufmann07}
T.~{Kaufmann}, L.~{Mayer}, J.~{Wadsley}, J.~{Stadel} and B.~{Moore}, {\em
  \mnras} {\bf 375}, 53 (2007).

\end{thebibliography}

\end{document}